\documentclass[fleqn,twoside]{article}
\usepackage{espcrc2}

\usepackage{graphicx}

\newcommand{\AmS}{{\protect\the\textfont2
  A\kern-.1667em\lower.5ex\hbox{M}\kern-.125emS}}

\title{A relation between CP violation of low energy and leptogenesis}

\author{G. C. Branco \address[MSCD]{Departamento de Fisica, Instituto
        Superior Tecnico,
        Av. Rovisco Pais, P-1049-001,Lisboa, Portugal\\},
        T. Morozumi \address[HIRO]{Graduate School of Science,
Hiroshima University, Higashi-Hiroshima, Japan, 739-8526,  \\ },%
 B. M. Nobre \addressmark[MSCD],
                and M. N. Rebelo \addressmark[MSCD] }
\begin{document}
\def\bea{\begin{eqnarray}}
\def\eea{\end{eqnarray}}
\def\nn{\nonumber}
\begin{abstract}
We discuss how CP violation generating lepton number asymmetry
can be related to CP violation in low energy.  
\vspace{1pc}
\end{abstract}

\maketitle

\section{Introduction}
CP violation at low energy is observed in K and B system.
In the future neutrino oscillation experiments , 
CP asymmetry of the neutrino oscillations $P(\nu_i  \rightarrow \nu_j) \ne 
P(\bar{\nu_i} \rightarrow \bar{\nu_j}) $ may be also measured.
Our question is how low energy CP violation measurements
are related to CP violation for Baryon number asymmetry. 
Fukugita and Yanagida proposed a scenario for Baryon number
asymmetry based on the seesaw model.\cite{see}
In their scenario, heavy Majorana neutrinos decays
give rise to lepton number asymmetry. The asymmetry is 
converted into Baryon number through spharelon process. 
\cite{Fukugita:1986hr} 
In this scenario, CP violating phases in the seesaw model
contribute to both CP violation at low energy and CP
violation for lepton number asymmetry. 
However, this correlation is not trivial.
This is partly because there are six 
independent CP violating phases in the seesaw model and 
the low energy CP violating observables are just three phases
of them.
It can be shown that three of the six 
CP violating sources may contribute to the
lepton number asymmetry.\cite{Endoh:2001hc}
We can ask the following question.
If CP asymmetry of 
the neutrino oscillations $P(\nu_i  \rightarrow \nu_j) \ne 
P(\bar{\nu_i} \rightarrow \bar{\nu_j}) $ is measured, what does it mean
about CP violation for leptogenesis. To answer to this question, 
we must first identify the number of the independent CP violating phases
and find which of them contributes to leptogenesis and
to neutrino oscillations.
The plan of my talk is following. We first review the counting of CP 
phases in the minimal seesaw model and explicitly construct a
parameterization. Then we identify the phases in leptogenesis
and CP violation in low energy. Finally we give a specific scenario 
in which both CP violating phenomena has a correlation.\\
\section{The number of independent CP violating phases in the minimal seesaw
model}
In the seesaw model, we have three sources for lepton mass terms;
namely, Charged Lepton Yukawa couplings
$m_l$, neutrino Yukawa couplings,
$m_D$ and Majorana mass terms $M_R$. 
\bea
{\cal L}=-[\overline{{\nu}_{L}^0} m_D N_{R}^0 +
\frac{1}{2} \bar{N_{R}^{0}}^{c}  M_R N_{R}^0+
\overline{l_L} m_l l_R]. 
\eea
By using a suitable basis transformation, we can choose the basis in which 
$m_l$ and $M_R$ are real diagonal. In this basis all the CP violation
is included into Dirac Yukawa term $m_D$. 
$m_D$ is $n_g \times n_g$ complex matrix, this contains 
${n_g}^2$ imaginary part. We can still absorb $n_g$ phases.
Therefore we obtain ${n_g}^2-n_g$ independent CP violating phases.
For $n_g=3$, we have six CP violation phases.
These six CP violating sources are identified 
in weak basis invariant way.\cite{bmnr}
The weak basis invariants are non-zero if CP is violated.
The six weak basis invariants are given as:
\bea
I_1&=&Im Tr[h H {M_R}^* h^* M_R], \nn \\
I_2&=&Im Tr[h H^2 {M_R}^* h^* M_R], \nn \\
I_3&=&Im Tr[h H^2 {M_R}^* h^* M_R H], \nn \\
I_4&=&Im Tr[\bar{h} H {M_R}^* \bar{h}^* M_R], \nn \\
I_5&=&Im Tr[\bar{h} H^2 {M_R}^* \bar{h}^* M_R], \nn \\
I_6&=&Im Tr[\bar{h} H^2 {M_R}^* \bar{h}^* M_R H]. 
\eea
where $h={m_D}^\dagger m_D$, $H={M_R}^{\dagger}M_R$,and  
$\bar{h}={m_D}^{\dagger} m_l {m_l}^{\dagger} m_D$.
\section{CP violating phases for leptogenesis}
CP violation for leptogenesis was computed in the
base in which the heavy Majorana mass matrix $M_R$ is real diagonal.
The lepton number asymmetry from the heavy Majorana particles
decay is proportional to the following combination.
\bea
Im[({m_D}^{\dagger} m_D)_{ij}]^2 (i \ne j).
\eea
This combination is independent of the left-handed rotation;
$ m_D \rightarrow g_L m_D$.
Therefore it is convenient to use the following parametrization.
\bea
m_D = U Y_{\Delta},
\eea 
where $U$ is a unitary matrix and $Y_{\Delta}$ is a triangular matrix.
The explicit parametrization for the unitary matrix is given as:
\bea
U=U_{23}({\theta_{23}}^{\prime}) U_{13} 
({\theta_{13}}^{\prime},\delta^{\prime}) U_{12}({\theta_{12}}^{\prime})
\times \nn \\ 
diag.(1, exp(i {\alpha^{\prime}}_1), exp(i {\alpha^{\prime}}_2)).
\eea
The triangular matrix is given as;
\bea
Y_{\Delta}=\left(\begin{array}{ccc} \nn 
                          Y_1 & 0 & 0 \\ 
                          Y_{21} & Y_2 & 0 \\
                          Y_{31} & Y_{32} & Y_3  
                          \end{array} \right). \\
\eea
Note that the diagonal elements $Y_1, Y_2$, and $Y_3$ are real.
We can easily confirm the decomposition
$
m_D = U Y_{\Delta}
$
counts correctly the independent parameters of $m_D$. 
$m_D$ (after removing three diagonal phases from the left)
has $6$ imaginary parameters and $9$ real parameters.
$Y_{\Delta}$ has $3$ imaginary parts and $6$ real parts
and $U$ has $3$ angles and $3$ phases. 
Using the decomposition, we can write the CP
violation relevant for leptogenesis as,
\bea
Im[({m_D}^{\dagger} m_D)_{ij}]^2 
=Im[({Y_{\Delta}}^{\dagger} Y_{\Delta})_{ij}]^2,
\quad i\ne j.
\eea
Therefore, CP violation phases for leptogenesis are
related to three phases, $argY_{ij}$ in $Y_{\Delta}$.
\section{The correlation between CP violation at low energy and leptogenesis}
Now we turn to CP violation in neutrino oscillation. The effective
mass matrix for light Majorana neutrinos in the seesaw model is given as;
\bea
m_{eff}=-m_D \frac{1}{M_R} {m_D}^T= -U Y_{\Delta} \frac{1}{M_R} {Y_{\Delta}}^T U^T.
\eea
Here the MNS matrix \cite{Maki:1962mu} $K$ is determined as:
\bea
 -K^{\dagger} m_D \frac{1}{M_R}{m_D}^T K^{*} = d.
\eea 
where $d=diag.(d_1,d_2,d_3)$, where $d_1,d_2$ and $d_3$ correspond to the
three mass eigenvalues for light neutrinos.
The low energy CP violation phases in $K$ are one Kobayashi-Maskawa type 
phase $\delta$ and two Majorana phases $\alpha_1, \alpha_2$. 
Our question is to what extent the CP violation phases in K is 
sensitive to  $argY_{ij}$ which contribute to the leptogenesis.
In general, the phases in $K$
are complicated functions of all of the six phases
$(\delta^{\prime}, {\alpha_1}^{\prime}, {\alpha_2}^{\prime},
arg(Y_{\Delta})_{ij})$. They also 
depend on heavy Majorana masses $M_R=diag.(M_1,M_2,M_3)$
and $|Y_{ij}|$.
To study the correlation between the low energy phases in
$K$ and phases for leptogenesis $Y_{\Delta}$, let us examine
the equation for diagonalization of the effective Majorana
mass matrix, $m_{eff}=-m_D \frac{1}{M_R} {m_D}^T$.
\bea
-K^{\dagger} U (Y_{\Delta} \frac{1}{M_R} {Y_{\Delta}}^T) U^T K^{*}=d
\eea
We can see that, in general, $K$ depends on the $Y_{\Delta}$.
Next we ask in what kind of situation, the correlation 
between a set of low energy phases  
$(\delta, \alpha_1, \alpha_2)$ and CP violating phases
for leptogenesis
$ (arg.Y_{21},arg.Y_{31},argY_{32})$ 
is weak and/or strong.
A key is the matrix $Y_{\Delta} \frac{1}{M_R} Y_{\Delta}^T$:
\begin{eqnarray}
&&\left(\begin{array}{ccc}
\frac{{Y_1}^2}{M_{1}}&
\frac{Y_1 Y_{21}}{M_{1}} & \frac{Y_1 Y_{31}}{M_{1}}
\\
\frac{Y_1 Y_{21}}{M_{1}}  & \frac{{Y_2}^2}{M_{2}}
+ \frac{{Y_{21}}^2}{M_{1}} &  \frac{Y_{21} Y_{31}}{M_{1}}+
\frac{Y_2 Y_{32}}{M_{2}}
  \\
\frac{Y_1 Y_{31}}{M_{1}} &
\frac{Y_{21} Y_{31}}{M_{1}}+
\frac{Y_2 Y_{32}}{M_{2}}
& \frac{{Y_3}^2}{M_{3}}+
\frac{{Y_{32}}^2}{M_{2}}+\frac{{Y_{31}}^2}{M_{1}}
\\
\end{array}\right). \nn \\
\end{eqnarray}
\begin{enumerate}
\item[A)]{ The case that the correlation is weak.\\
If $Y_{\Delta} \frac{1}{M_R} Y_{\Delta}^T$
are nearly diagonal, the neutrino mixings
must be accounted by $U$. Therefore, in this case,
\bea
K^{\dagger} U \sim 1 \rightarrow K \sim 
U({\alpha_i}^{\prime}, \delta^{\prime}).     
\eea
Such situation may be realized if
all $Y_{ij}$ are the same order and $M_1>>M_2>>M_3$.
If this is the case, the correlation between the low energy
phase and CP violation for leptogenesis may be weak.\cite{Endoh:2001hc}}
\item[B)]{
The case that the correlation is strong.\\
If $U \simeq 1$ and the elements of $Y_{\Delta} \frac{1}{M} Y_{\Delta}^T$
are nearly degenerate, we may except $K$ directly depends on the
phases of $Y_{\Delta}$. The case study with hypothesis
$Y_1=m_u,Y_2=m_c,Y_{3}=m_t$ and $U=1$ is done. See \cite{PEM}.
In this case, all the other parameters 
$|Y_{21}|, |Y_{31}|, |Y_{32}|$ $M_1,M_2, M_3$
$arg Y_{21}, arg Y_{31},arg Y_{32}$
can be determined from the low energy input
$d_1,d_2, d_3$ 
(light neutrino masses)
and MNS matrix: $\theta_{12}, \theta_{13}, 
\theta_{23}$ and  $\alpha_1, \alpha_2, \delta$.
It was shown that the hierarchy $M_1<<M_2<<M_3$ 
is required to obtain large mixing.}
\end{enumerate} 
\section{Conclusions}
\begin{enumerate}
\item{There are six independent CP violating phases in the seesaw model.
Among them, three contribute to lepton number asymmetry.}
\item{In the basis where $M_R$ and $m_l$ are real diagonal, all CP violating 
sources can be put into Yukawa term $m_D$.}
\item{A convenient parametrization of $m_D$ is proposed:
$m_D = U Y_{\Delta}$. The phases in $Y_{\Delta}$ exhausts the CP
violation for lepton number asymmetry, while there are the other
three phases in $U$.}
\item{MNS matrix $K$ is obtained from
$-K^{\dagger} (U Y_{\Delta} \frac{1}{M_R} Y_{\Delta}^T U^T)K^* =d$.
Therefore, in general, the CP violating phases in $K$ are 
sensitive for leptogenesis.}
\item{The cases with the correlation and without the correlation
are discussed qualitatively. In particular, the strong correlation
occurs if $U=1$. In this case, all the CP violating sources
of the standard model come from $Y_{\Delta}$ and low energy CP violation 
of neutrino sector can be related to leptogenesis phases.
There are the other cases the correlation does exist.\cite{bmnr} }
\end{enumerate}
{\bf Acknowledgement}\\
We would like to thank organizers of KEKTC5 and T. Endoh and 
A. Purwanto for discussion. The work of T. M. is supported by the
Grant-in-Aid for scientific research No.13640290 from the Ministry 
of education, science and culture of Japan. The work of BMN was
supported by Funda\c c\~ ao para a Ci\^ encia e a Tecnologia (FCT)
(Portugal) through fellowship SFRH/BD/995/2000; GCB, BMN and MNR
received partial support from FCT from Project CERN/P/FIS/40134/2000,
Project POCTI/36288/FIS/2000 and Project CERN/C/FIS/40139/2000. We
thank the CERN Theory Division for hospitality during the preparation
of Ref[4].

\end{document}